# Superconductivity at 5.5 K in $Nb_2PdSe_5$ compound

Reena Goyal[1,2], Govind Gupta[1], A.K. Srivastava[1] and V.P.S. Awana[1,*]

[1]*CSIR-National Physical Laboratory, Dr. K. S. Krishnan Marg, New Delhi-110012, India*
[2]*Academy of Scientific and Innovative Research, NPL, New Delhi-110012, India*

**Abstract**

We report superconductivity in as synthesized $Nb_2PdSe_5$, which is similar to recently discovered $Nb_2PdS_5$ compound having very high upper critical field, clearly above the Pauli paramagnetic limit [Sci. Rep. 3, 1446 (2013)]. A bulk polycrystalline $Nb_2PdSe_5$ sample is synthesized by solid state reaction route in phase pure structure. The structural characterization has been done by X ray diffraction, followed by Rietveld refinements, which revealed that $Nb_2PdSe_5$ sample is crystallized in monoclinic structure with in space group *C*2/*m*. Structural analysis revealed the formation of sharing of one dimensional $PdSe_2$ chains. Electrical and magnetic measurements confirmed superconductivity in $Nb_2PdSe_5$ compound at 5.5K. Detailed magneto-resistance results, exhibited the value of upper critical field to be around 8.2Tesla. The estimated $H_{c2}(0)$ is within Pauli Paramagnetic limit, which is unlike the $Nb_2PdS_5$.

**\*Corresponding Author**
Dr. V. P. S. Awana, Principal Scientist,
E-mail: awana@mail.npindia.org
Ph. +91-11-45609357, Fax-+91-11-45609310,
Homepage: awanavps.wenbs.com

**Introduction**

Superconductivity in $Nb_2PdS_5$ compound, which crystallizes in monoclinic space group C2/m, was first reported by Q. Zhang et. al. in 2013 [1]. The corresponding value of superconducting transition temperature ($T_c$) and upper critical field were found to be at 6K and above 20Tesla respectively [1]. The robustness of $Nb_2PdS_5$ superconductor against magnetic field aroused enormous interest in scientific community [1,2]. The upper critical field of $Nb_2PdS_5$ superconductor is clearly outside the Pauli Paramagnetic limit of $H_{c2}=1.84T_c$ [1,2]. The high value of upper critical field makes it distinct from various other superconductors discovered till now. The structure is layered and superconductivity supposedly resides in quasi one-dimensional PdS chains [1, 2]. Due to quasi one-dimensional structure, the nano wires of $Nb_2PdS_5$ could be grown easily [3]. Electronic structure calculation on $Nb_2PdS_5$ compound confirmed the multiband nature of superconductivity [1]. Specific heat measurements also provided the evidence for multiband nature of superconductivity in this class of materials [4]. Later on Park et. al. presented direct spectroscopic evidence for two band superconductivity in $Nb_2PdS_5$ compound [5]. N. Zhou et. al. elaborated the role of spin orbit coupling on upper critical field value by chemical doping on Pd chains with elements of different masses [6]. The ratio of $H_{c2}/T_c$ significantly increases by heavy element doping, while the same decreases with lighter element doping on Pd chains [6]. Niu et. al. studied the effect of Se doping on the S site in $Nb_2PdS_5$ superconductor [7]. They observed that superconductivity gradually suppresses for x<0.5 in $Nb_2Pd(S_{1-x}Se_x)_5$ and completely disappeared for x≥0.5[7]. Interestingly enough, Khim et. al. observed superconductivity in $Nb_2PdSe_5$ with $T_c$=5.5K [8]. In both reports [7,8], though the studied $Nb_2PdSe_5$ was crystallized in monoclinic structure within space group *C2/m*, i.e., similar to that as for $Nb_2PdS_5$ compound, but one is not [7], and other is [8] superconducting. More recently, Zhang et. al., reported superconductivity at around 2.5 K in another similar structure Selenide i.e., $Ta_2PdSe_5$ [9]. Clearly, contradictory reports [7,8] do exist on observation of superconductivity in $Nb_2PdSe_5$.

In short, robust superconductivity of layered transition metal based compounds i.e., $(Nb/Ta)_2Pd(S/Se/Te)_5$ with quasi one dimensional Pd(S/Se/Te) chains, is fast catching the attention of condensed matter physics community [1-12]. In this regards, as mentioned above, contradictory reports do exist [7, 8] on superconductivity in $Nb_2PdSe_5$. In current letter, we report the successful synthesis of phase pure superconducting $Nb_2PdSe_5$ compound by solid state synthesis method. Our results clearly demonstrate that the as synthesized nominal

Nb$_2$Pd$_{1.2}$Se$_5$ compound is a superconductor with transition temperature at 5.5K and an upper critical field value of 8.2 Tesla. Our results are in sharp contradiction to that as reported in ref. 7, but in agreement with ref.8. Summarily, we found that Nb$_2$PdSe$_5$ crystallizes in desired monoclinic structure within space group *C*2/*m* and is a 5.5K superconductor.

**Experimental Section**

The Nb$_2$Pd$_{1.2}$Se$_5$ sample is synthesized using standard solid state reaction route. The high purity element Nb (4N), Pd (4N) and Se (4N) were mixed in the stoichiometric ratios by continuous grinding in an argon filled glove box. The excess amount of nominal Pd is taken, to compensate possible loss of Pd, as being done earlier [4,12]. In, results section we now on words designate the compound as Nb$_2$PdS$_5$. The mixed powder was pelletized and sealed in an evacuated (10$^{-4}$ torr) quartz tube. The sealed tube was then kept in a furnace for sintering at 850$^o$C with a heating rate of 2$^o$C/min for 24 hours. Thus, obtained sample was once again pulverized, pelletized and sealed in an evacuated quartz tube. The sealed sample was kept in furnace again at 850$^o$C for another 24 hours and subsequently cooled to room temperature.

The structural characterization in terms of powder x-ray diffraction (XRD) is done on Rigaku X-ray diffractometer using the CuK$_α$ at 1.5414Å. Rietveld refinements of XRD data was carried out using Fullprof software. Detailed microstructural characterization of Nb$_2$PdSe$_5$ alloy was carried out using a high resolution transmission electron microscope (HRTEM, model: Tecnai G2-F30-STWIN assisted with the field emission gun for the electron source at an electron accelerating voltage of 300kV). The electronic binding energy of Nb$_2$PdSe$_5$ sample was examined using x-ray photoelectron spectroscopy (XPS) by Omicron multiprobe surface analysis system equipped with Mono Al-Kα excitation source (1486.7eV). The electrical and Magnetic measurements were performed on quantum design physical property measurement system (PPMS) equipped with superconducting magnet of 14Tesla. The magnetization measurements with temperature were taken in both field cooled (FC) and zero field cooled (ZFC) modes.

**Results and Discussion**

The XRD pattern of Nb$_2$PdSe$_5$ compound along with its Rietveld refinements is shown in Fig. 1. Rietveld refinement of XRD data is carried out using Fullprof software for determining crystal structure and information about lattice. Fig 1 suggests that the sample is single phase without any impurities within XRD resolution. The as synthesized Nb$_2$PdS$_5$ is crystallized in monoclinic structure with space group *C*2/*m*. The estimated lattice parameters are found to be *a*=12.769(5) Å, *b* =3.401(4) Å, *c* =15.476(2) Å and *β* =101.47(4). Apparently, the absence of any additional peaks, suggests the phase purity of as synthesized $Nb_2PdSe_5$ sample. The global fitness of XRD pattern is $\chi^2$=4.21. The refined structural parameters including various co-ordinate positions for different elements are shown in Table 1. The unit cell of the studied compound being drawn with help of Rietveld refined co-ordinate positions is shown in inset of Fig. 1. The details of unit cell are given along with transmission electron microscopy result.

Further, detailed micro structural characterization of $Nb_2PdSe_5$ sample was carried out using HRTEM. A uniform grain microstructure with a faceted contrast was delineated throughout in the microstructure (Fig. 2a). It was interesting to note that the grains were single crystalline in nature. The grey level contrast exhibiting fine grain-boundaries is presumably because of relatively different orientations between the adjacent grains (Fig. 2a). A set of corresponding selected area electron diffraction patterns (SAEDPs) recorded from different regions of the microstructure (Fig. 2a) shows the spotty array of diffraction patterns (Fig. 2b, and an inset in Fig. 2b), and inference about the single crystalline nature of the grains constituting the sample microstructure. A set of important planes with miller indices (hkl): 200, 0$\bar{2}$0, and 00$\bar{2}$ of a monoclinic crystal structure of $Nb_2PdSe_5$ (space group: *C*2/*m*, lattice constants: *a*=1.278nm, *b*=0.3391nm, *c*=1.541nm, *α*=*γ*=90º, and *β*=101.48º, reference: JCPDS card no. 811905) are marked over the electron diffraction patterns in Fig. 2b. At lattice scale, atomic arrangement of unit cell was resolved. It was interesting to observe that a stacking periodic array of a columnar structure was self repetitive throughout in the microstructure (Fig. 2c). A set of ten such periodic units are denoted by a set of arrows in fig. 2c. The separation between the two similar columnar units was measured to be about 1.32nm (Fig. 2c). It was further observed that in between these strong-contrast columns of atomic planes (marked by set of arrows), there was further a presence of atomic planes between these columns, which in turn revealed a complete networking of atomic planes in the unit cell of $Nb_2PdSe_5$. To understand the arrangement of atomic planes at lattice scale, here, the VESTA software was used to visualize the crystal structure of $Nb_2PdSe_5$ unit cell. The atomic coordinates of Nb, Pd, Se were taken from the Rietveld refinements [Table 1] for calculation purpose of $Nb_2PdSe_5$. Apparently, the crystal structure of $Nb_2PdSe_5$ consists of flat Nb-Pd-Se layers and a formation of linear chain of shared motif-$PdSe_2$ was observed along '*b*' direction of the axis of $Nb_2PdSe_5$ crystal lattice. In the structure calculation, the measured distance between Pd-Pd was found to be 1.276nm, as marked on Fig. 2d. The experimentally observed atomic scale arrangement with a periodic separation of about 1.32nm (Fig. 2c) was in corroboration with theoretically measured distance of same coordinates as 1.276 nm (Fig. 2d) of a repetition unit of PdSe$_2$.

Figure 3a represents the core level of Pd XPS spectra being split into two peaks corresponding to $3d_{5/2}$ and $3d_{3/2}$ at binding energy (B.E.) of 336.1eV and 341.4eV. A spin orbit splitting of 5.3eV is observed for Pd-3d region, which is in complete agreement with previous reports [13]. The $3d_{5/2}$ and $3d_{3/2}$ peaks were deconvoluted into four main peaks at 336.0eV, 337.2eV and 341.3eV, 342.5eV, respectively. The peaks at 336.0eV corresponds to the formation of Pd-Se bonds as there is shifting in binding energy towards higher side as compared to pure Pd(335.1eV) [13]. The peak at 337.2eV corresponds to the formation of palladium oxide on the surface [14, 15]. As shown in Fig. 3(b), the core level of Nb spectra splits into two peaks ($3d_{5/2}$ and $3d_{3/2}$), with spin orbit splitting of 2.8 eV [16]. The spectra were fitted by four main peaks with position at 203eV, 204eV, 205.8eV and 207eV, respectively. The peaks at 203eV and 205.8eV are assigned to the $Nb_{5/2}$ and $Nb_{3/2}$ of NbSe$_2$ [17]. The peaks at 204eV and 207eV assigned to the $Nb_{5/2}$ and $Nb_{3/2}$ peaks of Nb$_2$O$_5$ [17]. For selenium (Se), the 3d core level spectra were deconvoluted into six main peaks as clearly shown in Fig. 3(c). The B.E. shifts toward lower side in comparison to pure Se [17]. The shift in B.E. shows transfer of negative charge from selenium to niobium as well as palladium. The peaks at 52.8eV and 53.7eV (marked as □ in fig.) are assigned to $Se_{5/2}$ and $Se_{3/2}$ of Niobium diselenide [17]. On the other hand, the doublet peaks at 53.3eV and 54.2eV (marked as * in same Fig.) can be assigned to $Se_{5/2}$ and $Se_{3/2}$ peaks of hybridization of Selenium with Pd [18, 19]. The peaks at 54.9eV and 55.9eV are assigned to the formation of Selenium oxide [17].

Figure 4 shows the temperature dependence of resistivity for Nb$_2$PdSe$_5$ compound in temperature range of 2 to 300K. Clearly, the variation of resistivity at high temperature shows metallic behavior for as obtained polycrystalline Nb$_2$PdSe$_5$ sample. A sudden drop of resistivity can be clearly seen at low temperature of around 5.7K, corresponding to the occurrence of superconducting behavior in Nb$_2$PdSe$_5$ sample. The onset and offset ($\rho$=0) values of transition temperature are 5.7K and 5.5K respectively. Thus obtained superconducting transition width is 0.2K for studied Nb$_2$PdSe$_5$. In temperature range 6K to 70K i.e., just above the superconducting onset temperature, the normal state resistivity follows $\rho=\rho_0+AT^2$ criteria. The variation of resistivity with square of temperature is known to be strong evidence for the Fermi liquid behavior [1, 2, 12]. The solid red curve in Fig. 4 shows the resistivity fitting with above mentioned criteria. The residual resistivity ($\rho_0$) and temperature coefficient of electrical resistivity for $Nb_2PdSe_5$ sample are found to be 0.23m$\Omega$-cm and 8.93×10$^{-5}$m$\Omega$-cm-K$^{-2}$, respectively. The Residual resistivity ratio (*RRR*) given by the ratio of resistivity at room temperature to the resistivity at zero temperature is estimated to be 6.54, which is good enough for any polycrystalline bulk sample [12]. Inset of Fig. 4 depicts $\rho$-$T$ measurements in low temperature range from 2 to 7K under different applied magnetic fields varying from 0 to 8Tesla. Clearly, both the onset and offset of $T_c$ ($T_c^{onset}$ and $T_c^{\rho=0}$) shift toward the lower temperatures with increasing magnetic field. This behavior is expected for any typical type II superconductors. The upper critical field is an intrinsic property of type II superconductors, which can be extracted from resistive transition curve i.e., $\rho(T, H)$. Figure 5 represents the variation of upper critical field with temperature and is calculated from 90%, 50% and 10% of normal resistivity criteria. The value of upper critical field $H_{c2}(0)$ is estimated by using one band (*WHH*) equation i.e., $H_{c2}(0)= -0.69T_c(dH_{c2}/dT)_{T=T_c}$. Based on 90%, 50 % and 10 % of normal resistivity criteria the values of upper critical field are found to be 8.2Tesla, 7.0Tesla and 6.5Tesla respectively. The estimated upper critical filed value for $Nb_2PdSe_5$ superconductor is well within Pauli Paramagnetic limit, which is defined as $\mu_0H_p=1.84T_c$ [20]. The coherence length at T=0K was estimated using Ginzburg-Landau formula $\xi= (\Phi_0/2\pi\mu_oH_{c2}(0))^{1/2}$, where $\Phi_0=2.07 \times 10^{-7}$ Oe cm$^2$, calculated to be 6.34nm, 6.86nm and 7.12nm for $H_{c2}(0)$ at 90%, 50% and 10% of $\rho_n$ respectively.

The bulk superconducting behavior is confirmed by magnetization measurements on $Nb_2PdSe_5$ sample, which is given in Fig. 6. The magnetization measurements were carried in both field cooled (*FC*) and zero field cooled (*ZFC*) mode under applied *DC* field of 10Oe in low temperature range. In *ZFC* mode $Nb_2PdSe_5$ sample has been cooled in absence of applied magnetic field and in *FC* mode the sample is cooled with an applied magnetic field of 10Oe. A sharp diamagnetic signal is clearly seen at 5.5K, confirming superconductivity in $Nb_2PdSe_5$ at below 5.5K. This value of transition temperature matches with onset value of superconducting transition temperature obtained from electrical resistivity measurement data (Fig.4). It is clear from resistivity and magnetization measurements that the studied $Nb_2PdSe_5$ compound is a bulk superconductor at below 5.5K. Thus our results approve the findings of superconductivity in $Nb_2PdSe_5$ as reported in ref. 8 but in contrast to in ref.7.

Figure 7 represents isothermal magnetization (*MH*) plots at 2K for $Nb_2PdSe_5$ compound in its superconducting state. Apparently, as field increases magnetization decreases linearly up to $H_{c1}$=346Oe, suggesting perfect diamagnetic character of $Nb_2PdSe_5$ sample. We calculated the value of thermodynamic critical field at absolute zero i.e., $H_c(0)$ by using relation $H_c = (H_{c1}H_{c2})^{1/2}$, and the value of same comes out to be 5327.3 Oe. Using GL theory, the value of Ginzburg- Landau (GL) parameter κ is estimated from the relation between upper critical field and thermodynamic critical field i.e., $H_{c2}=2^{1/2}\kappa H_c$, and the resulting value is κ=10.8. In our case the value of GL parameter is greater than $(1/2)^{1/2}$, which implies that $Nb_2PdSe_5$ is a type II superconductor [21]. The penetration depth is calculated using the relation $\lambda(0) = \kappa\xi(0)$, and the value comes out to be 6.0 *nm*.

Summarily, we reported successful synthesis of phase pure $Nb_2PdSe_5$ crystallizing in desired monoclinic structure with in space group C2/m. Further both electrical and magnetic measurements showed that $Nb_2PdSe_5$ is a bulk 5.5 K superconductor.

**Acknowledgement**

We wish to acknowledge the support from the Director of NPL-CSIR India for his encouragement and providing research facilities. Reena Goyal, acknowledges the financial support from University Grant Commission-Senior research fellowship (UGC-SRF) and DAE-SRC outstanding investigator award scheme. We also thank S. P. Patnaik from Jawaharlal Nehru University-India and B. Tiwari from SVNIT-Surat-India for the magnetization measurements and XRD analysis.

**Table 1: Positional atomic data for $Nb_2PdSe_5$ sample.**

| Atom | x | y | z | site | Fractional occupancy |
|---|---|---|---|---|---|
| Nb1 | 0.0063(4) | 0.5000 | 0.1629 (4) | 4i | 1 |
| Nb2 | 0.1208(3) | 0.0000 | 0.3515 (3) | 4i | 1 |
| Pd1 | 0.0000 | 0.0000 | 0.0000 | 2a | 1/2 |
| Pd2 | 0.0000 | 0.0000 | 0.5000 | 2c | 1/2 |
| Se1 | 0.3131(4) | 0.0000 | 0.4918 (3) | 4i | 1 |
| Se2 | 0.1675(5) | 0.5000 | 0.2836 (4) | 4i | 1 |
| Se3 | 0.0215(3) | 0.0000 | 0.1641 (5) | 4i | 1 |
| Se4 | 0.5535(4) | 0.5000 | 0.1328 (4) | 4i | 1 |
| Se5 | 0.4879(4) | 0.0000 | 0.3199 (6) | 4i | 1 |

**Reference**

1. Q. Zhang, G. Li, D. Rhodes, A. Kiswandhi, T. Besara, B. Zeng, J. Sun, T. Siegrist, M. D. Johannes, and L. Balicas, Sci. Rep., **3**, 1446 (2013).
2. R. Jha, B. Tiwari, P. Rani, H. Kishan and V. P. S. Awana, J. Appl. Phys. **115**, 213903 (2014).
3. H. Yu, M. Zuo, L. Zhang, S. Tan, C. Zhang, and Y. Zhang, J. Am. Chem. Soc. **135**, 12987 (2013).
4. R. Goyal, B. Tiwari, R. Jha and V. P.S. Awana, J. Supercond. Nov. Mag. **28**, L1427-32 (2015).
5. E. Park, X. Lu, F. Ronning, J. D. Thompson, Q. Zhang, L. Balicas and T. Park, arxiv(2015)
6. N. Zhou, X. Xu, J. R. Wang, J. H. Yang, Y. K. Li, Y. Guo, W. Z. Yang, C. Q. Niu, B. Chen, C. Cao and J. Dai, Phys. Rev. B **90**, 094520 (2014).
7. C. Q. Niu, J. H. Yang, Y. K. Li, B. Chen, N. Zhou, J. Chen, L. L. Jiang, B. Chen, X. X. Yang, C. Cao, J. Dai and X. Xu, Phys. Rev. B **88**, 104507 (2013).
8. S. Khim, B. Lee, K. Y. Choi, B. G. Jeon, D. H. Jang, D. Patil, S. Patil, R. Kim, E. S. Choi, S. Lee, J. Yu, K. H. Kim, New J. Phys. **15**, 123031 (2013).
9. J Zhang, J. K. Dong, Y. Xu, J. Pan, L. P. He, L. J. Zhang and S. Y. Li, Supercond. Sci. Technol. **28**, 115015 (2015).
10. Y. Lu, T. Takayama, A. F. Bangura, Y. Katsura, D. Hashizume and H. Takagi, J. Phys. Soc. of Jpn. **83**, 023702 (2014).
11. W. H. Jiao, Z. T. Tang, Y. L. Sun, Y. Liu, Q. Tao, C. M. Feng, Y. W. Zeng, Z. A. Xu and G. H. Cao, J. Am. Chem. Soc. **136**, 1284 (2014).
12. R. Goyal, B. Tiwari, R. Jha and V. P.S. Awana, J. Supercond. Nov. Mag. **28**, L1195-98 (2015).
13. R. Bhatt, S. Bhattacharya, R. Basu, A. Singh, U. Deshpande, C. Surger, S. Basu, D. K. Aswal and S. K. Gupta, Thin solid films **539**, 41 (2013).
14. M. C. Militello, S. J. Simko, Surf. Sci. Spectra **3**, 4 (1997)
15. E. H. Voogt, A. J. M. Mens, O. L. J. Gijzeman and J. W. Geus, Surf. Sci. **350**, 21 (1996).
16. A. R. H. F. Ettema and C. Haas, J. Phys. Condens. Matter **5**, 3817-3826 (1993).
17. N. D. Boscher, C. J. Carmalt and I. P. Parkin, Eur. J. Inorg. Chem. 1255 (2006).
18. S. Kukunuri, M. R. Krishnan, S. Sampath, Phys. Chem. Chem. Phys. **17**, 23448 (2015).
19. Y. Zhang, Z. P. Qiao and X. M. Chen, J. Mater. Chem. **12**, 2747-2748 (2002).
20. A. M. Clogston, Phys. Rev. Letter **9**, 6 (1962).
21. B. Tiwari, R. Goyal, R. Jha, A. Dixit and V. P. S. Awana, Supercond. Sci. Technol. **28**, 055008 (2015).

**Figure Captions**

**Figure 1:** Room temperature XRD pattern of as synthesized $Nb_2PdSe_5$ sample along with its Rietveld refinements, inset shows the schematic unit cell of the same.

**Figure 2:** HRTEM micrographs of $Nb_2PdSe_5$ alloy showing (a) bright field electron micrograph, (b) corresponding selected area electron diffraction patterns, (c) atomic scale image of a columnar structure, and (d) a calculated schematic of a unit cell of $Nb_2dSe_5$ crystal structure.

**Figure 3:** The experimental and curve fitted XPS spectra of (a) $Pd_{3d}$, (b) $Nb_{3d}$ and (c) $Se_{3d}$ for $Nb_2PdSe_5$ sample.

**Figure 4:** Temperature dependence of electrical resistivity for $Nb_2PdSe_5$. Inset shows the same under different applied magnetic fields in superconducting region.

**Figure 5:** Upper critical field of $Nb_2PdSe_5$ sample as a function of temperature. Solid lines represent linear extrapolation of experimental data fitted with WHH equation.

**Figure 6:** Temperature dependence of DC magnetization for $Nb_2PdSe_5$ sample in both ZFC and FC mode under applied magnetic field of 10Oe

**Figure 7:** Isothermal magnetization (MH) plots for $Nb_2PdSe_5$ superconductor at 2K.

Figure 1

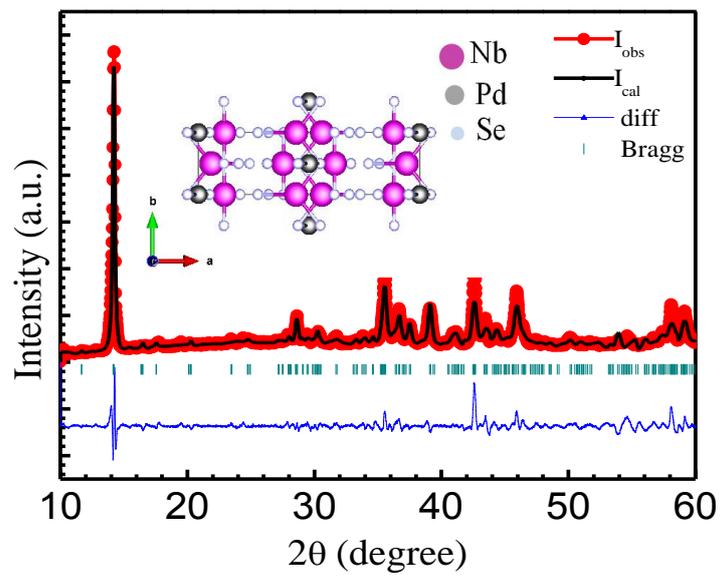

Figure 2

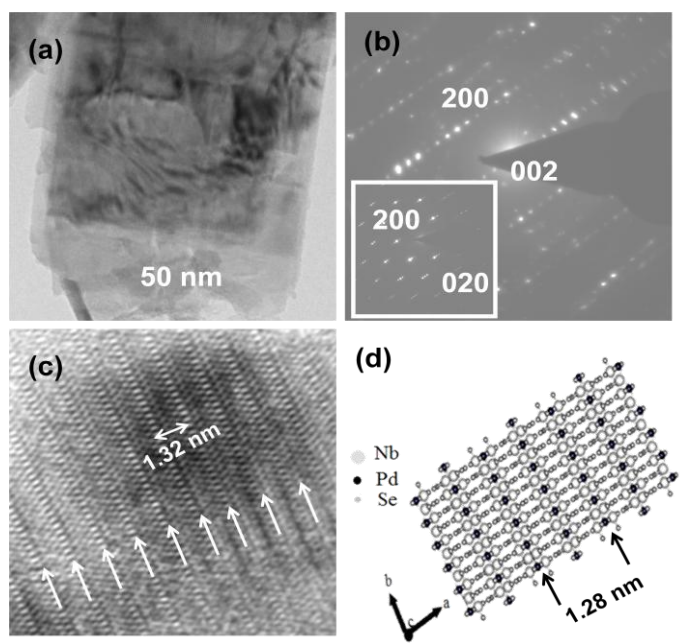

Figure 3

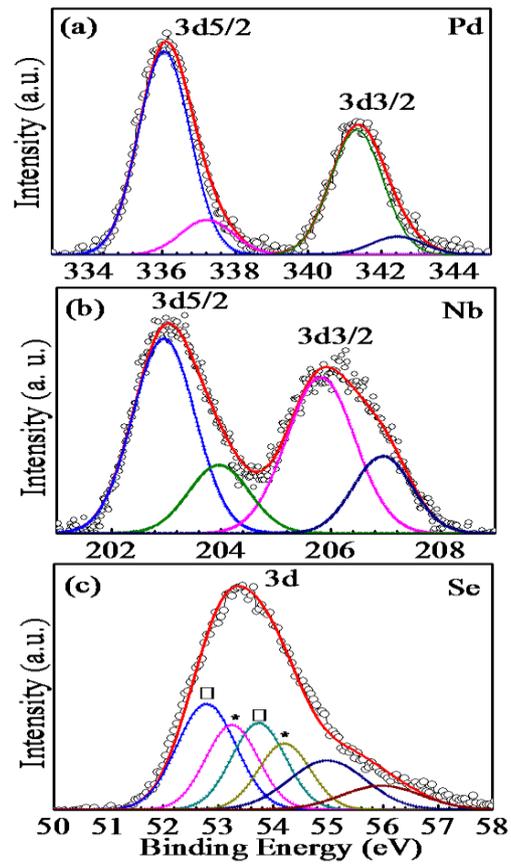

Figure 4

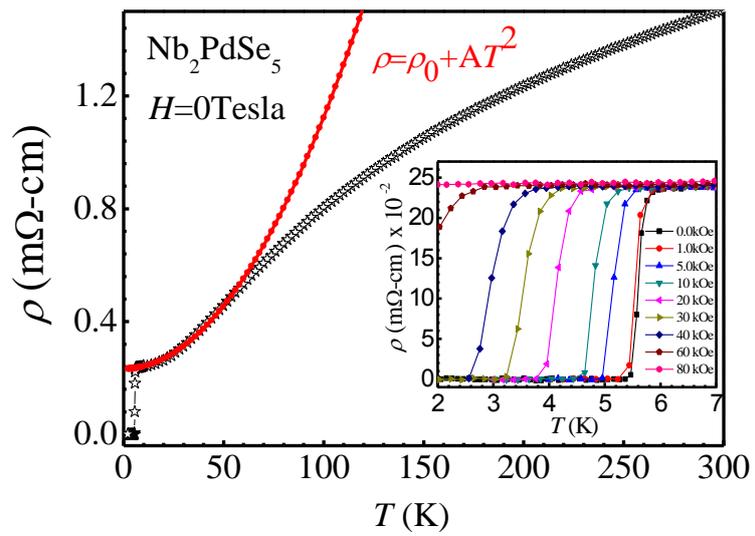

Figure 5

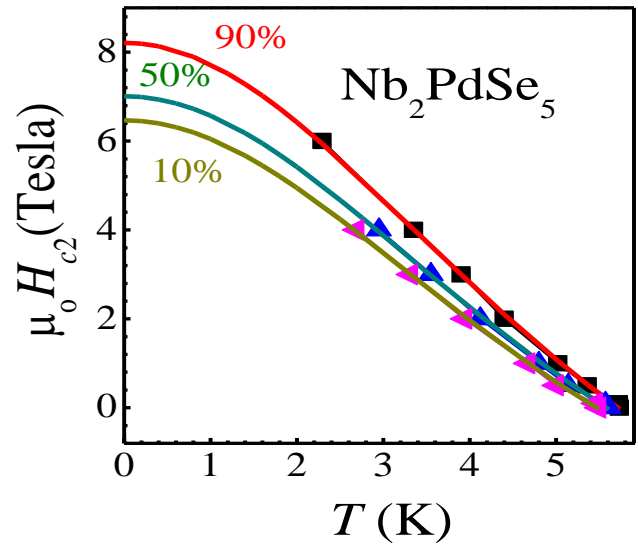

Figure 6

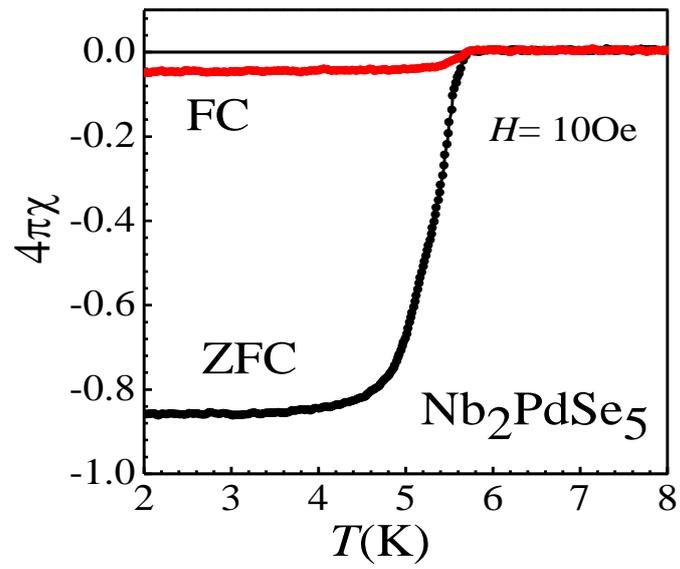

Figure 7

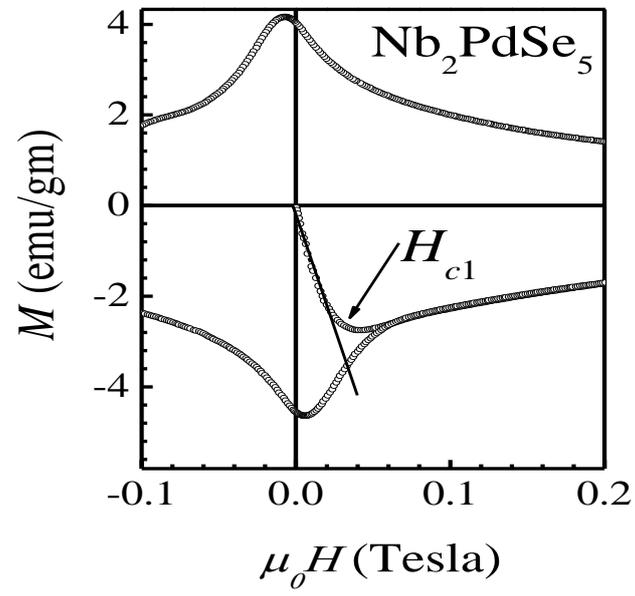